\documentclass[doublecol,linenumbers]{epl2}
\usepackage{amsthm,amsmath,amsfonts,dsfont,upgreek} %bbm,bbold,,txfonts,mathbbol
\usepackage{amssymb,epsfig,setspace}
\usepackage{graphicx}
\usepackage[english]{babel}
\usepackage{verbatim}
\usepackage{bm}% bold math

\begin{document}

\hyphenation{ano-ther ge-ne-ra-te dif-fe-rent know-le-d-ge po-ly-no-mi-al}
\hyphenation{me-di-um or-tho-go-nal as-su-ming pri-mi-ti-ve pe-ri-o-di-ci-ty}
\hyphenation{mul-ti-p-le-sca-t-te-ri-ng i-te-ra-ti-ng e-q-ua-ti-on}
\hyphenation{wa-ves di-men-si-o-nal ge-ne-ral the-o-ry sca-t-te-ri-ng}
\hyphenation{di-f-fe-r-ent tra-je-c-to-ries e-le-c-tro-ma-g-ne-tic pho-to-nic}
\hyphenation{Ray-le-i-gh di-n-ger Kra-jew-ska Wal-czak Ham-bur-ger Ad-di-ti-o-nal-ly}
\hyphenation{Kon-ver-genz-the-o-rie ori-gi-nal in-vi-si-b-le cha-rac-te-ri-zed}
\hyphenation{Ne-ver-the-less sa-tu-ra-te Ene-r-gy sa-ti-s-fy le-vels re-s-pec-ti-ve pro-pe-r-ty}
\hyphenation{dif-fe-rent no-men-cla-tu-re re-gar-ding di-n-ger en-e-r-gies sa-ti-s-fies mul-ti-va-lu-ed di-f-fe-ren-tial}

\title{On the Heisenberg condition in the presence of redundant poles of the S-matrix}

\author{
Alexander Moroz\inst{1} \and Andrey E. Miroshnichenko\inst{2} %\thanks{http://www.wave-scattering.com}
%\\
%Wave-scattering.com
} 
\institute{
\inst{1}Wave-scattering.com\\
\inst{2}School of Engineering and Information Technology,
University of New South Wales Canberra
Northcott Drive, Campbell, ACT 2600, Australia
%%\\
%%andrey.miroshnichenko@unsw.edu.au 
}

%%% a maximum of three PACS 
%%% \pacs{02.30.Ik}{Integrable systems}
%%\pacs{03.65.Ge, 02.30.Ik}

%%\begin{abstract}\noindent
\abstract{
For the same potential as originally studied by Ma [Phys. Rev. {\bf 71}, 195 (1947)] we obtain 
analytic expressions for the Jost functions and the residui of the S-matrix of both (i) redundant 
poles and (ii) the poles corresponding to true bound states. 
This enables us to demonstrate that the Heisenberg condition is valid in spite of the presence of 
redundant poles and singular behaviour of the S-matrix for $k\to \infty$. 
In addition, we analytically determine the overall contribution of redundant poles to the asymptotic 
completeness relation, provided that the residuum theorem can be applied. 
The origin of redundant poles and zeros is shown to be related to peculiarities of 
analytic continuation of a parameter of two linearly independent analytic functions. 
}
%%\end{abstract}
%%% https://authors.epletters.net/4DCGI/_authors&how_to_prepare/-1/0/0/0/0/0/836663154
%%% http://www.charactercountonline.com/

\maketitle 

%%% units; chemical formulae; the differential d and the mathematical 
%%% functions, cos, sin, exp, det, ker, ln; tr, 
%%% Tr (for traces); Im and Re (for imaginary and real); 
%%% imaginary unit i; Euler constant e.
%%% should be typeset in roman
%%%
%%% The maximum length for manuscripts is 6 printed journal pages

%%%% $m$, $\hbar$ is the reduced Planck constant, $E$, $V(x)$

%%%% fictitious, illusory, z-variable, 
%%%% confluent hypergeometric function ${}_0F_1(1-{\rm i}\rho,-\alpha^2/4)$
%%%% surprizing absence of exact results for Bessel functions of general complex order

\section{Introduction}
\label{sc:intr}
%%%%%%%%%%%%%%%%%%%%%%%%%%%%%%%%%%%%%
One of the classical results of scattering theory is that the S-matrix may possess
{\em redundant} zeros \cite{Ma1,Ma2,Ma3}, {\em i.e.} {\em not} corresponding to any physical state. This bears
important consequences for relating analytic properties the S-matrix to physical states.
In particular, the presence of redundant zeros implied that the S-matrix need not always satisfy 
a general condition of Heisenberg \cite{Ma2,tH,BPS}. The issue is thus of fundamental importance for 
the S-matrix theory. The result was first obtained by Ma \cite{Ma1,Ma2,Ma3}
when analyzing the $s$-wave S-matrix for an attractive exponentially decaying potential \cite[pp. 110-111]{BB}
\begin{equation}
V(r)=-V_0 {\rm e}^{-r/a},
\label{edp}
\end{equation}
where $V_0>0$ and $a>0$ are positive constants and $r$ is radial distance. 
Ma's \cite{Ma1,Ma2,Ma3} results
inspired and motivated many authors, such as in now classical refs. \cite{tH,BPS,Jst47,Brg,vanK,Ps,LR}.

In what follows we will demonstrate that the concept of redundant zeros, 
at least for the potential (\ref{edp}), is far from being fully understood. We confirm the existence 
of those redundant zeros, but not for a reason given by Ma \cite{Ma1,Ma2,Ma3}.
One can construct a proper regular solution that does not vanish 
identically at the position of original redundant zeros of this example
by a proper choice of the basis of linearly independent solutions.
The crux of the appearance of redundant zeros lies in that analytic continuation of a parameter of two linearly 
independent functions may result in linearly dependent 
functions at an infinite discrete set of isolated points of the parameter complex plane.

At the same time, especially in connection with non-Hermitian scattering Hamiltonians \cite{SBK},
one witnesses a recent revival of interest in the analytic structure of the S-matrix
leading to a number of surprising real applications \cite{Aliprl102}.
In this regard, the potential (\ref{edp}) has an immense pedagogical value.
Indeed it satisfies conditions \cite[eqs. (12.20) and (12.21)]{RN}
sufficient to prove analyticity of the S-matrix, but merely in a strip around the real axis in 
the complex plane of momentum $k$ \cite[p. 352]{RN}. 
Nevertheless, the $s$-wave S-matrix for the potential (\ref{edp}) 
can be determined analytically in the whole $k$ complex plane, what the classical 
monograph \cite{RN} surprisingly never mentions. Therefore, the physically relevant potential (\ref{edp}) 
provides one with a unique window of opportunity to illustrate the validity of general theorems
and examine the properties of the S-matrix.
It will be shown here that there is a number of important lessons 
to be learned from this example.

The outline of the letter is as follows. We first review Ma's analysis \cite{Ma1,Ma2,Ma3} 
and highlight its weak points. Afterwards, a rigorous canonical analysis of the $s$-wave S-matrix 
is provided along the lines of monograph \cite{RN}.
We obtain analytic expressions for the Jost functions and the residui of the S-matrix (\ref{MaS}) of both 
(i) redundant poles [eq. (\ref{rprs}]
and (ii) the poles corresponding to true bound states [eqs. (\ref{MaSr})-(\ref{Jres})]. In addition, we 
analytically determine the overall contribution of redundant poles [eq. (\ref{tcrp})]
in the asymptotic completeness relation [eq. (\ref{hgc})].
We demonstrate that the Heisenberg condition [eqs. (\ref{hu15}), (\ref{hu15sf})] is valid despite
the presence of infinite number of redundant poles on the physical sheet and singular behavior of the
S-matrix for $k\to\infty$. In a remarkable twist of history, the model which was the very origin 
for doubting the validity of the Heisenberg condition (\ref{hu15}) will be shown to be
a perfect example of just the opposite.

\section{Bethe and Bacher analysis}
\label{sc:bban}
%%%%%%%%%%%%%%%%%%%%%%%%%%%%%%%%%%%%%
For the sake of conciseness, let us summarize the essential steps of Bethe and Bacher \cite{BB}
in analyzing the potential (\ref{edp}). With the substitution $\psi(r)=u(r)/r$ one has
\begin{eqnarray}
\psi''(r)+ \frac{2}{r}\, \psi'(r) &=& \frac{u''(r)}{r},
\nonumber
\end{eqnarray}
where prime denotes derivative with respect to the function argument.
Therefore, the radial $s$-wave equation takes on the form \cite[p. 108, eq. (33)]{BB}
\begin{equation}
u''(r)+[k^2 + U_0 {\rm e}^{-r/a}] u=0,~~ k^2=\frac{2m}{\hbar^2}\, E,~~ U_0=\frac{2m}{\hbar^2}\, V_0.
\label{edper}
\end{equation}
Ma \cite{Ma1,Ma2,Ma3} took for the general solution of (\ref{edper})
a linear combination of Bessel function of complex order ${\rm i}\rho$, where
$\rho=2ak$ is a dimensionless momentum parameter,
\begin{eqnarray}
u(r) &=& c_1 J_{{\rm i}\rho}\left(2a{\rm e}^{-r/(2a)}\sqrt{U_0} \right)
\nonumber\\
&& +\,  c_2 J_{-{\rm i}\rho}\left(2a{\rm e}^{-r/(2a)} \sqrt{U_0} \right),
\label{gedps}
\end{eqnarray}
with $c_1$ and $c_2$ being arbitrary integration constants.
Indeed, with {\em e.g.} $c_1=1$, $c_2=0$, and $\sigma={\rm e}^{-r/a}$, $x=2a\sqrt{U_0\sigma}$, one has
\begin{eqnarray}
u'(r) &=& -\sqrt{U_0 \sigma} J_{{\rm i}\rho}'(x),
\nonumber\\
u''(r) &=& U_0\sigma J_{{\rm i}\rho}''(x) +\frac{\sqrt{U_0\sigma}}{2a}\, J_{{\rm i}\rho}'(x).
\nonumber
\end{eqnarray}
When substituting back into (\ref{edper}), one arrives at 
\begin{equation}
U_0\sigma J_{{\rm i}\rho}''(x) +\frac{\sqrt{U_0\sigma}}{2a}\,
  J_{{\rm i}\rho}'(x) + [U_0\sigma + k^2]J_{{\rm i}\rho}(x)=0.
\nonumber
\end{equation}
After multiplication by $4a^2$,
\begin{equation}
 x^2 J_{{\rm i}\rho}''(x) +x\, J_{{\rm i}\rho}'(x) + [x^2-({\rm i}\rho)^2] J_{{\rm i}\rho}(x)=0.
 \label{dbfe}
\end{equation}
The latter can be compared with the defining equation of Bessel functions of order
$\nu$ (cf. (9.1.1) of \cite{AS}),
\begin{equation}
z^2 w''(z)+z w'(z) + (z^2-\nu^2) w(z)=0.
\nonumber
\end{equation}

\section{Ma's derivation of the $s$-wave S-matrix}
\label{sc:masm}
%%%%%%%%%%%%%%%%%%%%%%%%%%%%%%%%%%%%%%%%%%%%%%%%%%%
The solution (\ref{gedps}) {\em regular} ({\em i.e.} vanishing)
at the origin is
\begin{eqnarray}
u(r) &=& C\left[ J_{-{\rm i}\rho}(\alpha) J_{{\rm i}\rho}\left(\alpha {\rm e}^{-r/(2a)} \right)\right.
\nonumber\\
&&\left. - J_{{\rm i}\rho}(\alpha) J_{-{\rm i}\rho}\left(\alpha {\rm e}^{-r/(2a)} \right)\right],
\label{gedprs}
\end{eqnarray}
where $\alpha=2a\sqrt{U_0}=\lim_{r\to 0} x(r)$ and 
$C$ is a normalization constant that is to ensure $u'(0)=1$ \cite[eq. (12.2)]{RN}.

According to (9.1.7) of \cite{AS}, or \cite[eq. (10.7.3)]{Ol}, one has in the limit $z\to 0$,
\begin{equation}
J_\nu(z)\sim \frac{z^\nu}{2^\nu \Gamma(\nu+1)}\qquad (\nu\ne -1, -2, -3, \ldots).
\label{as9.1.7}
\end{equation}
Hence the regular solution (\ref{gedprs}) has the following
asymptotic behaviour in the limit $r\to \infty$:
\begin{eqnarray}
u(r) &\sim& C J_{-{\rm i}\rho}(\alpha) \frac{\alpha^{{\rm i}\rho}}{2^{{\rm i}\rho} \Gamma(1+{\rm i}\rho)}
{\rm e}^{-{\rm i}kr}
\nonumber\\
&& 
- C J_{{\rm i}\rho}(\alpha) \frac{2^{{\rm i}\rho}}{\alpha^{{\rm i}\rho} \Gamma(1-{\rm i}\rho)} {\rm e}^{{\rm i}kr}.
\label{gedprsa}
\end{eqnarray}
The latter expression determines the S-matrix as follows \cite[eq. (14)]{Ma2}
\begin{equation}
S(k)= \frac{J_{{\rm i}\rho}(\alpha) \Gamma(1+{\rm i}\rho)}{J_{-{\rm i}\rho}(\alpha) \Gamma(1-{\rm i}\rho)}
\left(\frac{\alpha}{2}\right)^{-2{\rm i}\rho},
\label{MaS}
\end{equation}
where $k$ dependence here enters through dimensionless momentum parameter $\rho=2ak$. 
It follows straightforwardly that $S(k)$ vanishes when either 
(i) $J_{{\rm i}\rho}(\alpha)=0$ or 
(ii) $\Gamma(1-{\rm i}\rho)$ is infinite \cite{Ma1,Ma2,Gmf}. 
Given that $J_{{\rm i}\rho}(\alpha)$ has no zeros in the lower 
half of complex $k$-plane except those on the imaginary axis, one arrives at the 
implicit condition on allowable $k$-values \cite[eq. (42f)]{BB}, \cite[eq. (15)]{Ma2}
\begin{equation}
J_{|\rho|}(\alpha)=0,
\label{BBc}
\end{equation}
yielding the physical bound states \cite{BB,Ma1,Ma2}.

The poles of $\Gamma(1-{\rm i}\rho)$ occur for any ${\rm i}\rho=n\in\mathbb{N}_+$. 
Those are the only poles of $\Gamma(1-{\rm i}\rho)$, and all those poles are simple \cite{AS}.
On using the relationship for $z\in\mathbb{C}$ \cite[eq. (9.1.5)]{AS}:
\begin{equation}
J_{-n}(z)=(-1)^n J_n(z),
\label{jpmm}
\end{equation}
eq. (\ref{gedprs}) can be recast for ${\rm i}\rho=\eta$, $\eta\in\mathbb{Z}$ as
\begin{eqnarray}
u(r) &=& C\left[ J_{-\eta}(\alpha) J_\eta(x) -
 J_\eta(\alpha) J_{-\eta}(x)\right]
\nonumber\\
&=& (-1)^n C \left[ J_\eta(\alpha) J_\eta(x)-J_\eta(\alpha) J_\eta(x)\right] \equiv 0.
\nonumber
%%\label{gedprsr}
\end{eqnarray}
In particular, the regular solution (\ref{gedprs}) corresponding to
any of the poles ${\rm i}\rho=n\in\mathbb{N}_+$ of $\Gamma(1-{\rm i}\rho)$ vanishes identically 
in the lower half $k$ complex plane for any $k=-k_n$, $k_n={\rm i}n/(2a)$, $n\ge 1$ \cite{Ma2}.
The latter prompted Ma \cite{Ma1,Ma2} to call those zeros of the S-matrix {\em redundant},
because they did not correspond to any regular solution. 

From a historical perspective it is interesting that whereas the 
{\em redundant} zeros of the S-matrix (\ref{MaS}) received a lot of attention, 
nobody \cite{Ma2,Ma3,tH,BPS,Jst47,Brg} seemed to be
worried that a formal limit ${\rm i}\rho\to -n$, $n\ge 1$ ($k\to k_n$) yields 
a {\em redundant} pole singularity of the S-matrix (\ref{MaS}) 
on the physical sheet of energy in the upper half of complex $k$-plane.

\section{Pitfalls in Ma's derivation}
\label{sc:mshc}
%%%%%%%%%%%%%%%%%%%%%%%%%%%%%%%%%%%%%%%
A closer inspection of the above peculiarities of the S-matrix (\ref{MaS}) shows that they both 
can be traced to the fact that analytic expression (\ref{MaS}) was rigorously derived
only with the exclusion of the integer points ${\rm i}\rho\in\mathbb{Z}$ 
[any negative integer $\nu\in\mathbb{Z}_-$ was excluded from the asymptotic (\ref{as9.1.7})].
As if this was not enough, the Bessel functions $J_{\pm n}(x)$ become linearly 
{\em dependent} for any ${\rm i}\rho\in\mathbb{Z}\ne 0$ [cf. eq. (\ref{jpmm})].
Indeed, the Wronskian of $J_{\pm {\rm i}\rho}$ is proportional to $\sin ({\rm i}\rho \pi)$ 
\cite[eq. (9.1.15)]{AS}, \cite[eq. (10.5.1)]{Ol},
\begin{equation}
 W_x\{J_{{\rm i}\rho}(x),J_{-{\rm i}\rho}(x)\}=-\frac{2 \sin ({\rm i}\rho \pi)}{\pi x},
\label{wj+-}
\end{equation}
and hence vanishes whenever ${\rm i}\rho\in\mathbb{Z}$.
In the special case ${\rm i}\rho=-n$, $n\in\mathbb{N}$, one finds on combining eqs. (\ref{as9.1.7}), (\ref{jpmm}):
\begin{equation}
J_{-n}(z)\sim (-1)^n \frac{z^n}{2^n \Gamma(n+1)},\qquad n\in\mathbb{N}.
\label{as9.1.7m}
\end{equation}
Note that when repeating the steps that led us the S-matrix (\ref{MaS}), but now with 
asymptotic (\ref{as9.1.7m}) for ${\rm i}\rho\in\mathbb{Z}\ne 0$, one would find, formally, that 
(i) $\Gamma(-{\rm i}\rho+1)\to \Gamma(n+1)$ for ${\rm i}\rho=n$ in the denominator, whereas (ii) 
$\Gamma({\rm i}\rho+1)\to \Gamma(n+1)$ for ${\rm i}\rho=-n$ in the numerator of the S-matrix (\ref{MaS}). 
Crucially, this would prevent
(i) any redundant zero for $k=-k_n$ and 
(ii) any redundant pole for $k=k_n$ of the S-matrix
in the $k$ complex plane. All that before taking into account 
that the basis $\{J_{{\rm i}\rho}(z),J_{-{\rm i}\rho}(z)\}$ of solutions 
of eq. (\ref{dbfe}) collapses into linearly {\em dependent} 
solutions for any ${\rm i}\rho\in\mathbb{Z}$. 
(For $\rho=0$ the two Bessel functions degenerate into a single one.)

\section{A rigorous analysis of the $s$-wave S-matrix}
\label{sc:nsm}
%%%%%%%%%%%%%%%%%%%%%%%%%%%%%%%%%%%%%%%%%%%%%%%%%%%%%%%%%%%
In what follows we shall prove rigorously the statements made in the abstract.
To this end, we use the basis of solutions of eq. (\ref{dbfe}) remaining linearly independent for any parameter value. 
We follow the usual practice ({\em e.g.} when treating electromagnetic scattering 
from dielectric objects \cite{BH,AMap}) and choose a Bessel function of the second kind $Y_{{\rm i}\rho}$ 
(also known as Weber's or Neumann's function) \cite{AS} 
in place of $J_{-{\rm i}\rho}$ as a second linearly independent solution to $J_{{\rm i}\rho}$ of eq. (\ref{dbfe}). 
Indeed, in contrast to the pair $\{J_{{\rm i}\rho}(z),J_{-{\rm i}\rho}(z)\}$, 
the pair $\{J_{{\rm i}\rho}(z),Y_{{\rm i}\rho}(z)\}$ yields always two linearly independent solutions 
of eq. (\ref{dbfe}) and its Wronskian is never zero [cf. eqs. (9.1.15) and (9.1.16) of ref. \cite{AS}]. 
The {\em regular} solution of (\ref{gedprs}) vanishing at the origin $r=0$ becomes
in the notation of ref. \cite{RN}
\begin{eqnarray}
\varphi(r) &=& C\left[ Y_{{\rm i}\rho}(\alpha) J_{{\rm i}\rho}\left(\alpha {\rm e}^{-r/(2a)} \right)\right.
\nonumber\\
&& \left. - J_{{\rm i}\rho}(\alpha) Y_{{\rm i}\rho}\left(\alpha {\rm e}^{-r/(2a)} \right)\right],
\label{gedprsy}
\end{eqnarray}
where $C=\pi a$ ensures normalization $\varphi'(0)=1$ (see supplementary material).
Unlike Ma's regular solution (\ref{gedprs}), one notices immediately that 
$\varphi(r)$ does {\em not} vanish identically for any ${\rm i}\rho=n\in\mathbb{N}_+$ \cite[eq. (26)]{tH}.

$Y_{{\rm i}\rho}(z)$ is related to $J_{{\rm i}\rho}(z)$ by \cite[eq. (9.1.2)]{AS}, \cite[eq. (10.2.3)]{Ol}
\begin{eqnarray}
Y_{{\rm i}\rho}(z) &=&\frac{J_{{\rm i}\rho}(z)\cos ({\rm i}\rho \pi) -J_{-{\rm i}\rho}(z)}{\sin ({\rm i}\rho \pi) }
\nonumber\\
&=& \frac{J_{{\rm i}\rho}(z)\cosh (\rho \pi) - J_{-{\rm i}\rho}(z)}{{\rm i}\sinh (\rho \pi)}\cdot
\label{ynujnu}
\end{eqnarray}
When ${\rm i}\rho$ is an integer the right-hand side is replaced by
its limiting value \cite[eq. (10.2.4)]{Ol}.
Given the above relation, the usual irregular solution for $k\in\mathbb{R}$ is 
proportional to $J_{-{\rm i}\rho}(x)$,
\begin{eqnarray}
\lefteqn{
f_+(k,r)= \Gamma(1-{\rm i}\rho)\left(\frac{\alpha}{2}\right)^{{\rm i}\rho}\times
}
\nonumber\\
 && \left[\cosh (\rho\pi) J_{{\rm i}\rho}(x)
-{\rm i}\sinh (\rho\pi) Y_{{\rm i}\rho}(x)\right]
\nonumber
\nonumber\\
 &&= \Gamma(1-{\rm i}\rho)\left(\frac{\alpha}{2}\right)^{{\rm i}\rho} J_{-{\rm i}\rho}(x).
\label{f+}
\end{eqnarray}
The asymptotic (\ref{as9.1.7}) implies for Im $k\ge 0$ [Re $(-{\rm i}\rho)\ge 0$]
\begin{equation}
f_+(k,r) \sim {\rm e}^{{\rm i}kr},
\label{Jsas}
\end{equation}
showing the characteristic outgoing spherical wave behaviour of $f_+(k,r)$ for $r\to\infty$, $k\in\mathbb{R}$,
and yields $f_+(k,r)$ as exponentially decreasing for $r\to\infty$, Im $k> 0$, in accordance
with general theorems \cite[Sec. 12.1.4]{RN}. 
Given analyticity of $\Gamma(1-{\rm i}\rho)$ and $J_{-{\rm i}\rho}$ \cite{Ann}, one can easily verify
$f_+(k,r)$ to be for each $r$ an analytic function
of $k$ regular for Im $k>0$ and continuous with a continuous $k$ derivative in
the region Im $k\ge 0$. The second linearly independent irregular solution 
$f_-(k,r)=f_+(k{\rm e}^{{\rm i}\pi},r)\propto J_{{\rm i}\rho}(x)$ 
(assuming analytic continuation via the upper half plane)
is uniquely determined by the boundary condition $f_-(k,r) \sim {\rm e}^{-{\rm i}kr}$ for $r\to\infty$.

The Jost function is \cite[Sec. 12]{RN}
\begin{eqnarray}
\lefteqn{
{\cal F}_+(k):=W_r\{f_+,\varphi\}= C \Gamma(1-{\rm i}\rho)\left(\frac{\alpha}{2}\right)^{{\rm i}\rho}\times
}
\nonumber\\
 && \left[- \cosh (\rho\pi) J_{{\rm i}\rho}(\alpha) 
+{\rm i}\sinh (\rho\pi)Y_{{\rm i}\rho}(\alpha)\right] W_r\{J_{{\rm i}\rho},Y_{{\rm i}\rho}\}
\nonumber\\
 &&
= - C J_{-{\rm i}\rho}(\alpha) \Gamma(1-{\rm i}\rho)\left(\frac{\alpha}{2}\right)^{{\rm i}\rho}
\frac{2}{\pi \alpha {\rm e}^{-r/(2a)}} \left(\alpha {\rm e}^{-r/(2a)} \right)'
\nonumber\\
 &&
= J_{-{\rm i}\rho}(\alpha) \Gamma(1-{\rm i}\rho)\left(\frac{\alpha}{2}\right)^{{\rm i}\rho}
\nonumber\\
 &&
= {}_0F_1(1-{\rm i}\rho,-\alpha^2/4),
\label{Fkd}
\end{eqnarray}
where the respective $W_r\{.\, ,.\}$ and prime denote the Wronskian 
(cf. \cite[eq. (9.1.16)]{AS}, \cite[eq. (10.5.2)]{Ol}) 
and derivative with respect to $r$, and ${}_0F_1$ is the confluent hypergeometric function.

The complementary Jost function ${\cal F}_-(k):=W_r\{f_-,\varphi\}$ is obtained by replacing 
${\rm i}\rho\to -{\rm i}\rho$ in the above expression for ${\cal F}_+$.
The ratio $S(k)={\cal F}_-(k)/{\cal F}_+(k)$ (cf. \cite[eq. (12.71)]{RN})
then reproduces the S-matrix (\ref{MaS}).
${\cal F}_+(k)$ is analytic (without any singularity) on the physical sheet (Im $k\ge 0$), 
where it can have only zeros - in our case for any $J_{-{\rm i}\rho}(\alpha)=0$ corresponding 
to the poles of the S-matrix (\ref{MaS}). 
For Im $k >0$, $f_+(k,r)$ is exponentially decreasing at infinite $r$. 
If ${\cal F}_+(k_l) = 0$ there (which is in our case equivalent to $J_{|\rho|}(\alpha)=0$, 
and hence to the condition (\ref{BBc}) yielding the physical bound states \cite{BB,Ma1,Ma2}),
the solutions $f_+$ and $\varphi$ are necessarily multiples of one another \cite[eq. (12.49)]{RN}.
Therefore $\varphi$ has to be also regular at the spatial infinity, 
and thus a regular square-integrable wave function, with $k_l$ corresponding to an eigenvalue,
{\em i.e.} a bound state. This can be explicitly confirmed in present case 
for bound states $k_l ={\rm i}\kappa_l$, $\kappa_l >0$, on positive imaginary axis.
Whenever $J_{2a\kappa_l}(\alpha)=0$ for $\kappa_l >0$,
eq. (\ref{ynujnu}) substituted into (\ref{gedprsy}) yields
\begin{eqnarray}
\varphi_l (r) &=& \pi a J_{-2a\kappa_l}(\alpha) \left[
 \cot (-2a\kappa_l \pi) J_{-2a\kappa_l}(x)
\right.
\nonumber\\
&& \left. - Y_{-2a\kappa_l}(x)\right] 
= -\frac{\pi a}{\sin(2a\kappa_l\pi)}
\nonumber\\
& \times& J_{-2a\kappa_l}(\alpha) J_{2a\kappa_l}\left(\alpha {\rm e}^{-r/(2a)} \right).
\label{gedprsyb}
\end{eqnarray}
Now exponential decrease for $r\to\infty$ is obvious from the asymptotic (\ref{as9.1.7}).
For the potential (\ref{edp}), the number of bound states in the $s$-channel can be estimated 
to be \cite{Brgb} [cf. eq. (\ref{edper})]
\begin{equation}
n_l \le \int_0^\infty |V(r)|\, r dr=a^2 U_0=\alpha^2/4.
\label{co1.2}
\end{equation}
For an illustration, a number of bound states as a function of $\alpha$ 
is shown in fig. \ref{fgHsnb}. In particular, for $\alpha<2$ the S-matrix (\ref{MaS}) 
has infinite number of redundant poles on the physical sheet without a single bound state.

The present $f_\pm(k,r)$ satisfy all the classical requirements \cite{RN}.
The usual analytic connection between the positive and negative real
$k$ axis, $f_-(k,r)=f_+(k{\rm e}^{{\rm i}\pi},r)$,
together with the boundary condition satisfied by $f_\pm$ leads to
$S(-k) = S^*(k) = S^{-1}(k)$ for any $k\in\mathbb{R}$ \cite[eq. (12.74)]{RN}.
For general $k\in\mathbb{C}$ one has (cf. \cite[eqs. (12.24a), (12.32a)]{RN})
\begin{equation}
S^*(k^*) = S^{-1}(k),
\label{rn12.32}
\end{equation}
which can be readily verified for the S-matrix (\ref{MaS})
(all the special functions involved there satisfy the Schwarz reflection principle
$F(\bar{z})=\overline{F(z)}$ in variable $z={\rm i}\rho$ for $\alpha\in\mathbb{R}$). 
Hence each pole of S on the first physical sheet of energy 
(Im $k> 0$) corresponds to a zero of S on the second sheet (Im $k< 0$), 
and vice versa \cite[Sec. 12.1.4]{RN}.

\section{Points ${\rm i}\rho\in\mathbb{Z}$}
\label{sc:nsmp}
%%%%%%%%%%%%%%%%%%%%%%%%%%%%%%%%%%%%%%%%%%%
For the future discussion it is important to notice that eqs. (\ref{f+}), (\ref{Fkd})
imply {\em factorization} of $f_\pm(k,r)$ as
\begin{equation}
f_\pm(k,r)= \frac{ {\cal F}_\pm (k)}{J_{\mp {\rm i}\rho}(\alpha)}\, J_{\mp {\rm i}\rho}(x),
\label{fctrz}
\end{equation}
where the first factor including the Jost function, ${\cal F}_\pm$, is only a function of $k$, 
and only the second factor, $J_{\mp {\rm i}\rho}(x)$, depends on both $k$ and $r$.
In virtue of (\ref{Fkd}), the first factor is finite for any $J_{\mp {\rm i}\rho}(\alpha)=0$.
 
Obviously one can get rid of the pair $\{J_{{\rm i}\rho}(x),J_{-{\rm i}\rho}(x)\}$
in the regular solution $\varphi(r)$ but not in the irregular solutions $f_\pm$.
The collapse of the pair $\{J_{{\rm i}\rho}(x),J_{-{\rm i}\rho}(x)\}$ of solutions 
of eq. (\ref{dbfe}) into linearly {\em dependent} solutions for any ${\rm i}\rho\in\mathbb{Z}$ brings about some
interesting peculiarities.
Let us ignore for a while the first $k$-dependent prefactors in (\ref{fctrz}).
Then $f_-(k,r)$, which is typically exponentially {\em increasing} on the physical 
sheet as $r\to \infty$, would become suddenly exponentially {\em decreasing} for 
$r\to \infty$ for any ${\rm i}\rho\in\mathbb{Z}_-$, {\em i.e.} $k=k_n$, $n\ge 1$, on the physical sheet, 
very much the same as $f_+(k,r)$. Similarly, $f_+(k,r)$, which is expected to be 
exponentially {\em increasing} on the second sheet for $r\to \infty$,
would become suddenly exponentially {\em decreasing} in the limit 
for any ${\rm i}\rho\in\mathbb{Z}_+$, or $k=-k_n$ on the 2nd sheet, very much the same as $f_-(k,r)$.
The role of the $k$-dependent prefactors ${\cal F}_\pm$ is to hide such an 
``embarrassing'' behaviour by causing
the respective irregular solutions $f_\pm(k,r)$ to become {\em singular} at the 
incriminating points ({\em i.e.} $f_-(k,r)$ at $k=k_n$, and $f_+(k,r)$ at $k=-k_n$).
Note in passing that although $f_+(k,r)$ ($f_-(k,r)$) is, for each $r$, an analytic function
of $k$ regular for Im $k>0$ (Im $k<0$) and continuous with a continuous $k$ derivative in
the region Im $k\ge 0$ (Im $k\le 0$), this no longer holds for Im $k<0$ (Im $k>0$). 

The singular prefactors ensure that, in spite of the linear dependency 
of the pair $\{J_{{\rm i}\rho}(x),J_{-{\rm i}\rho}(x)\}$ for any ${\rm i}\rho\to\mathbb{Z}$, the identity 
$W\{f_+,f_-\}=-2{\rm i}k$ \cite[eq. (12.27)]{RN} is nevertheless 
preserved. Indeed (\ref{wj+-}) implies for ${\rm i}\rho\to n\in\mathbb{N}_+$
\begin{equation}
 W_x\{J_{-{\rm i}\rho}(x),J_{{\rm i}\rho}(x)\}\sim \frac{2(-1)^n ({\rm i}\rho-n)}{x}\to 0.
\nonumber %%%\label{wj+-l}
\end{equation}
At the same time the residues of $\Gamma(1-{\rm i}\rho)$ in the ${\rm i}\rho$ variable at those points are:
\begin{equation}
\left. \operatorname{Res} \Gamma(1-{\rm i}\rho)\right|_{{\rm i}\rho=-n} =\frac{(-1)^{n}}{(n-1)!}\cdot
\label{gmres}
\end{equation}
Therefore, in the limit ${\rm i}\rho\to n\in\mathbb{N}_+$,
\begin{eqnarray}
W_r\{f_+,f_-\} &=&
\Gamma(1-{\rm i}\rho)\Gamma(1+{\rm i}\rho) W_r \{J_{-{\rm i}\rho}(x),J_{{\rm i}\rho}(x)\}
\nonumber\\
&=& \frac{2 n}{z} \left(\alpha {\rm e}^{-r/(2a)} \right)' = - \frac{n}{a}= - 2{\rm i}k,
\label{wf+-l}
\end{eqnarray}
where $k=-k_n$. Analogously for ${\rm i}\rho\to n\in\mathbb{N}_-$, or $k=k_n$.

In virtue of the condition (\ref{BBc}), any zero of the Bessel function of integer order $n\ge 1$ 
corresponds to the value of $\alpha$ at which a true bound state coincides with the redundant pole at $k_n$.
The latter does not alter the above singular behaviour of $f_\pm(k,r)$.
In virtue of eq. (\ref{jpmm}), the ratio of Bessel functions of integer order $J_{\pm n}(\alpha)$ 
in (\ref{MaS}) reduces to $(-1)^n$, irrespective of their argument. Therefore, the S-matrix
maintains its zero at $k=-k_n$ and pole (but now physical one) at $k=k_n$.
The sole change is that, in these exceptional cases, the Jost function ${\cal F}_+(k)$ attains
a finite nonsingular value at $k=-k_n$ (cf. (\ref{gmres}) 
and \cite[eq. (9.1.66)]{AS}, \cite[eq. (10.15.3)]{Ol}).

\section{Heisenberg condition}
\label{sc:hgc}
%%%%%%%%%%%%%%%%%%%%%%%%%%%%%%%%%%%%%%%%%
The {\em completeness} relation involving continuous and 
discrete spectrum yields \cite[eq. (12.128)]{RN} 
\begin{equation}
\frac{2}{\pi} \int_0^\infty \frac{\phi^*(k,r)\phi(k,r')}{|{\cal F}_+(k)|^2}\,k^2 dk
+
\sum_l \frac{\phi_l^*(r)\phi_l(r')}{N_l^2} =\delta_{r,r'},
\label{hu11}
\end{equation}
where $\delta_{r,r'}=\delta(r-r')$ and  $N_l^2=\int_0^\infty [\varphi_l(r)]^2\, dr$ 
is the $l$th bound state (\ref{gedprsyb}) 
normalization constant.
In the limit $r\to\infty$ one gets from \cite[eqs. (12.35), (12.71), (12.73)]{RN} 
for $k>0$:
\begin{equation}
\phi(k,r)\sim \frac{{\cal F}_+(k) {\rm e}^{{\rm i}\delta(k)}}{k}\, \sin[kr+\delta(k)],
\label{rscs}
\end{equation}
where $\delta(k)$ is the scattering phase-shift \cite[eq. (12.95)]{RN}.
Given that $S(k)={\rm e}^{2{\rm i}\delta(k)}$ for $k\in\mathbb{R}$,
one can on using asymptotic form (\ref{rscs}) of 
regular solutions for $r,r'\gg 1$ in the completeness relation (\ref{hu11}) 
arrive at \cite[eq. (6)]{Ma2}, \cite[eq. (1.2)]{BPS}, \cite[eq. (13)]{Hu}
\begin{equation}
\int_{-\infty}^\infty S(k) {\rm e}^{{\rm i}k(r+r')}\, dk = \sum_l |C_l|^2 {\rm e}^{-|k_l|(r+r')},
\label{hgc}
\end{equation}
where 
\begin{equation}
|C_l|^2= \frac{2\pi}{\Gamma^2(1+2a\kappa_l)\left(\int_0^\infty J_{2a\kappa_l}^2(x)\, dr\right)}
 \left(\frac{\alpha}{2}\right)^{4a\kappa_l}
\label{C_l}
\end{equation}
[cf. the asymptotic of $\phi_l$ given by (\ref{gedprsyb})
that follows from (\ref{as9.1.7})].
Under the condition that the integral over the real axis can be closed by infinite
semicircle $\gamma$ in the upper half $k$-plane, {\em i.e.}
\begin{equation}
\oint_\gamma S(k) {\rm e}^{{\rm i}k(r+r')}\, dk = 0,
\label{hu16}
\end{equation}
one arrives at the correspondence between the poles of the S-matrix and bound states,
\begin{equation}
\oint_{k=k_l} S(k)\, dk = |C_l|^2 >0,
\label{hu15}
\end{equation}
where $\oint_{k=k_l}$ stands for integration along a contour encircling a single isolated bound state.
This correspondence is known as the {\em Heisenberg condition} \cite{Ma2,tH,BPS}.

In what follows we shall first determine the overall contribution of redundant poles 
to the integral on the lhs of (\ref{hgc}) as the sum over all residui.
On making use of eqs. (\ref{jpmm}) and (\ref{gmres}) in (\ref{MaS}), one finds
the following residuum in the ${\rm i}\rho$ variable for any $k_n$,
\begin{eqnarray}
\mbox{Res } S(k_n) &=& \frac{(-1)^n J_{n}(\alpha) }{J_{n}(\alpha) \Gamma(n+1)}
\left(\frac{\alpha}{2}\right)^{2n}\, \mbox{Res } \Gamma(-n+1)
\nonumber\\
&=& \frac{(-1)^n}{n!} \left(\frac{\alpha}{2}\right)^{2n}\, \frac{(-1)^{n}}{(n-1)!}
\nonumber\\
&=& \frac{1}{n!(n-1)!} \left(\frac{\alpha}{2}\right)^{2n}.
\nonumber %%\label{MaS}
\end{eqnarray}
When converting from ${\rm i}\rho$ to $k$ as independent variable, 
the left hand side of (\ref{hu15}) for $n$th redundant pole yields a positive number
as in the case of true bound states,
\begin{equation}
2\pi {\rm i}\, \mbox{Res}_k S(k_n) 
 = \frac{\pi}{a} \frac{1}{n!(n-1)!} \left(\frac{\alpha}{2}\right)^{2n}> 0.
\label{rprs}
\end{equation}
The overall contribution of redundant poles to the integral 
on the lhs of (\ref{hgc}) is
\begin{eqnarray}
\lefteqn{
2\pi {\rm i} \sum_{n=1}^\infty \mbox{Res}_k S(k_n) {\rm e}^{-n(r+r')/(2a)}
}
\nonumber\\
&& 
=
\frac{\pi}{a} \sum_{n=1}^\infty \frac{1}{n!(n-1)!} 
  \left[\frac{\alpha}{2}\, {\rm e}^{-(r+r')/(4a)}\right]^{2n}
\nonumber\\
&& 
= \frac{\pi}{a}\,q I_1(q)>0,
\label{tcrp}
\end{eqnarray}
where $q=(\alpha/2) {\rm e}^{-(r+r')/(4a)}$ and $I_1$ is the modified Bessel function of the first 
kind \cite[eq. (9.6.10)]{AS}, \cite[eq. (10.25.2)]{Ol}.

Redundant poles of $S(k)$ on the physical sheet seemingly spoil
the Heisenberg condition. Indeed, for any redundant pole, the left-hand side of (\ref{hu15}) is positive,
whereas the right-hand side is zero ($C_l\ne 0$ only for physical bound states). 
Surprisingly enough, the Heisenberg condition (\ref{hu15}) remains valid
despite the presence of infinite number of redundant poles on the physical sheet. 
%%%%%%%%%%
\begin{figure}
\begin{center}
\includegraphics[width=8cm,clip=0,angle=0]{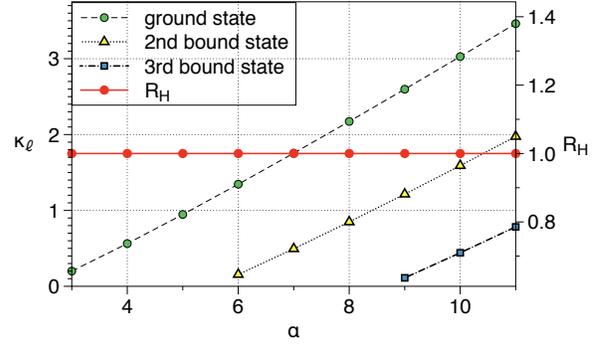}
\end{center}
\caption{$\kappa_l >0$ of corresponding bound states $k_l ={\rm i}\kappa_l$ on the positive imaginary axis
as a function of $\alpha$ for $a=1$. The number of bound states can be 
estimated from the upper  bound (\ref{co1.2}). 
For each bound state the ratio $R_H$ of lhs to rhs of the Heisenberg condition (\ref{hu15})
is shown. We get the ratio $R_H=1$ in machine precision on $21$ decimal places.
In the case of several bound states for a given $\alpha$, the corresponding ratios overlay 
each other and obviously cannot be distinguished by the naked eye.
}
\label{fgHsnb}
\end{figure}
%%%%%%%%%%
In fig. \ref{fgHsnb} the ratio $R_H$ of the lhs to rhs of the Heisenberg condition (\ref{hu15}) is plotted for each bound state
as obtained by Wolfram's Mathematica. The lhs was determined by complex integration, whereas
$|C_l|^2$ on the rhs was determined numerically from (\ref{C_l}). Without exception, 
the ratio $R_H=1$ in machine precision on $21$ decimal places.
In the case of several bound states for a given $\alpha$, the corresponding ratios necessary 
overlay each other and obviously cannot be distinguished
by naked eye. 

The above numerical evidence begs a deeper look at the Heisenberg condition (\ref{hu15}).
The lhs of the Heisenberg condition (\ref{hu15}) can be evaluated by the residuum theorem
as $2\pi {\rm i}\, \mbox{Res}_k S(k_l)$. At a bound state at $k=k_l=i\kappa_l$ ($\rho_l=2ak_l$) one has 
\begin{eqnarray}
\mbox{Res}_k S(k_l)&=& \frac{J_{{\rm i}\rho_l}(\alpha) \Gamma(1+{\rm i}\rho_l)}{\Gamma(1-{\rm i}\rho_l)}
\left(\frac{\alpha}{2}\right)^{-2{\rm i}\rho_l} 
\nonumber\\
&& \times \mbox{Res}_k \frac{1}{J_{-{\rm i}\rho_l}(\alpha)}\cdot
\label{MaSr}
\end{eqnarray}
Because $J_{-{\rm i}\rho}(\alpha)$ is holomorphic in its order $-{\rm i}\rho$ \cite{Ann},
the residuum of the pole term $1/J_{-{\rm i}\rho}(\alpha)$ in (\ref{MaSr})
can be obtained as
\begin{equation}
\mbox{Res}_k \frac{1}{J_{-{\rm i}\rho}(\alpha)} =\frac{{\rm i}}{2a}
          \left. \frac{1}{\partial_\nu J_\nu (\alpha)}\right|_{\nu=-{\rm i}\rho_l=2a\kappa_l}.
\label{Jres}
\end{equation}
At the same time, the integral in the denominator of $|C_l|^2$ in eq. (\ref{C_l}) 
can be performed analytically. On making the substitution $dx=-(1/2a)\, x\, dr$ one finds
\cite[\& 5$\cdot$11(15)]{Wat}, \cite[eq. (1.13.2.6) for $a=1$]{PBM2}
\begin{eqnarray}
\lefteqn{
\int_0^\infty J_{2a\kappa_l}^2(x)\, dr =
 2a\,\int_0^\alpha J_{2a\kappa_l}^2(x)\, \frac{dx}{x}
 }
\nonumber\\
&& = \left.  \frac{\alpha}{2\kappa_l} J_{2a\kappa_l+1}(\alpha)\, \frac{\partial J_\nu(\alpha)}{\partial\nu}
\right|_{\nu=2a\kappa_l}. 
\label{C_lfr}
\end{eqnarray}
On substituting (\ref{C_lfr}) into (\ref{C_l}) and, on combining with eqs. (\ref{MaSr}), (\ref{Jres}),
the Heisenberg condition (\ref{hu15}) reduces to
\begin{equation}
\frac{4a\kappa_l}{\alpha}
\frac{1}{\Gamma(1+2a\kappa_l)J_{2a\kappa_l+1}(\alpha)}= - J_{-2a\kappa_l}(\alpha) \Gamma(1-2a\kappa_l).
\nonumber %%\label{hu15s}
\end{equation}
Given the {\em reflection} formula,
\begin{equation}
\Gamma(1+2a\kappa_l)\Gamma(1-2a\kappa_l)=\frac{2a\kappa_l\pi}{\sin(2a\kappa_l\pi)},
\nonumber %%\label{rflid}
\end{equation}
the Heisenberg condition (\ref{hu15}) eventually becomes (see supplementary material for more detail)
\begin{equation}
J_{-2a\kappa_l}(\alpha) J_{2a\kappa_l+1}(\alpha) = -\frac{2\sin(2a\kappa_l\pi)}{\pi\alpha}\cdot
\label{hu15sf}
\end{equation}
We have checked numerically that (\ref{hu15sf}) is valid at the zeros $\kappa_l>0$ of $J_{2a\kappa_l}(\alpha)$.
However eq. (\ref{hu15sf}) is {\em not} an identity and does not hold for an arbitrary $\kappa_l$.

\section{Discussion}
\label{sc:dis}
%%%%%%%%%%%%%%%%%%%%%%%
The exponentially decaying potential (\ref{edp}) with physical applications \cite{BB} 
was shown to provide a beautiful laboratory for studying properties of the S-matrix.
Despite innocuous Schr\"odinger equation (\ref{edper}) which does not show 
any peculiarity for $k=\pm k_n$, the resulting S-matrix (\ref{MaS}) exhibits unexpected rich behaviour.
In particular, the S-matrix (\ref{MaS}) has always infinite number of redundant poles 
on the physical sheet even if there is not a single bound state for $\alpha<2$.

There is a number of important lessons to be learned from this example.
Despite all odds, the Heisenberg condition (\ref{hu15}), which was shown to reduce to 
analytic relation (\ref{hu15sf}), holds (cf. fig. \ref{fgHsnb}). 
Consequently, in virtue of the overall contribution of redundant poles (\ref{tcrp}),
the equality in (\ref{hgc}) cannot be preserved if one had attempted to perform 
the integral on the lhs of (\ref{hgc}) by closing the integration 
over the real axis in (\ref{hgc}) by infinite semicircle $\gamma$ in the upper half $k$-plane 
and replace it by the sum of residui of all enclosed poles.
This points to a problematic behaviour of the S-matrix (\ref{MaS}) for $k\to \infty$.
Indeed, because the number of redundant poles $k_n\to{\rm i}\infty$ on the physical sheet is {\em infinite}, 
the S-matrix (\ref{MaS}) {\em cannot} be analytic at infinity. 
Because $S(k_n)=\infty$, this also applies to the limit on the sequence $k_n\to\infty$.
On the other hand, one finds that the S-matrix (\ref{MaS}) has the following 
limit on the sequence $k={\rm i}(n+\tfrac12)/(2a)$ (${\rm i}\rho=-n-\tfrac12$), $n\to\infty$, 
(see supplementary material)
\begin{eqnarray}
S(k) &\sim& \frac{2(2n)!!}{(2n-1)!! \sqrt{2\pi(2n+1)} }\cdot
\label{MaSln}
\end{eqnarray}
In fact, the S-matrix is known to have in general an {\em essential singularity} for infinite $k$ 
and the class of cases for which the S-matrix is analytic for $k=\infty$ is very limited \cite{Ps}.
Therefore, the integral over the real axis in (\ref{hgc}) cannot be closed by infinite
semicircle $\gamma$ in the upper half $k$-plane. If it could be somehow closed,
one cannot exclude that a contribution of the contour integral (\ref{hu16}) will
cancel the contribution of (\ref{tcrp}) of redundant poles, thereby restoring
the asymptotic completeness relation (\ref{hgc}).
Alas, surprising absence of exact results for Bessel functions of general complex order \cite{AS,Ol}
provides a true obstacle in full analytic analysis of that issue. 

Another valid point is that the use of asymptotic form (\ref{rscs}) of 
regular solutions in the completeness relation (\ref{hu11}) imply that
the relation (\ref{hgc}) is not a rigorous identity. It involves only leading asymptotic
terms of regular solutions for $r,r'\to\infty$ leaving behind subleading terms,
which may also contribute exponentially small terms in (\ref{hgc}).

We have shown that the Heisenberg condition (\ref{hu15}) can be reduced down to eq. (\ref{hu15sf}).
Interestingly, we could find the above relation (\ref{hu15sf}) neither in tables \cite{AS,Ol} nor in monograph \cite{Wat}.
Therefore, we can at present confirm its validity at the zeros $\kappa_l$ of $J_{2a\kappa_l}(\alpha)$ only numerically.

Our analysis sheds new light on the redundant poles and zeros of the S-matrix (\ref{MaS})
of Ma \cite{Ma1,Ma2}. We have their following interpretation: 
the redundant poles (zeros) correspond to the points where the irregular solution $f_-(k,r)$ 
($f_+(k,r)$) and the Jost function ${\cal F}_-(k)$ (${\cal F}_+(k)$) become {\em singular}
in the upper (lower) half complex $k$-plane. The origin of those singularities is that in analytic continuation 
of a parameter of two linearly {\em independent} 
functions one cannot exclude that one ends up with linearly {\em dependent} 
functions at a discrete set (which can be infinite) of isolated points in the parameter complex plane.
In view of the {\em factorization} (\ref{fctrz}) of each $f_\pm(k,r)$,
the above singularities of $f_\pm$ and ${\cal F}_\pm$ are 
in fact indispensable for preserving the fundamental identity 
$W\{f_+,f_-\}=-2{\rm i}k$ \cite[eq. (12.27)]{RN}. Without the above singular
behaviour of $f_\pm$ and ${\cal F}_\pm$ one would in fact face discontinuities 
of $W\{f_+,f_-\}=-2{\rm i}k$ for any ${\rm i}\rho\to\mathbb{Z}$.
The above singular behaviour is also essential in preserving 
the classical statement that if $f_+$ and $f_-$ exist, 
they are linearly {\em independent}, except when $k=0$ \cite[p. 336]{RN}, {\em i.e.} 
at the point where $W\{f_+,f_-\}=-2{\rm i}k\to 0$. Without the above singular behaviour,
$f_+$ and $f_-$ would exist and be linearly {\em dependent} for any $k=\pm k_n\ne 0$.

\section{Conclusions}
\label{sc:con}
%%%%%%%%%%%%%%%%%%%%%%%%%%%%%%%%%%%%%%%%%%%%%%%%%%%%%%
For the same exponentially decaying potential (\ref{edp})
as originally studied by Ma \cite{Ma1,Ma2} we have obtained analytic expressions for the 
Jost functions and the residui of the S-matrix (\ref{MaS}) of both 
(i) redundant poles [eq. (\ref{rprs})] and 
(ii) the poles corresponding to true bound states [eqs. (\ref{MaSr})-(\ref{Jres})]. 
This enabled us to demonstrate that the Heisenberg condition (\ref{hu15}),
which was reduced down to analytic relation (\ref{hu15sf}), is valid despite 
the presence of infinite number of redundant poles on the physical sheet and singular behaviour of 
the S-matrix (\ref{MaS}) for $k\to \infty$. In a remarkable twist of history, the model which was 
the very origin for doubting the validity of the Heisenberg condition (\ref{hu15}) is now
a perfect example of just the opposite. We have analytically determined also the 
overall contribution of redundant poles [eq. (\ref{tcrp})] to the integral in (\ref{hgc}), 
provided that the contribution can be evaluated by the residuum theorem. The origin of redundant 
poles was shown to be related to peculiarities of analytic continuation 
of a parameter of two linearly {\em independent} analytic functions. 

Given that redundant poles and zeros occur already for such a simple model
is strong indication that they could be omnipresent.
Currently one can immediately conclude that the appearance of poles of ${\cal F}_+(k)$, 
and of the S-matrix, at $k_n= {\rm i}n/2a$ for positive integers $n$ is a general feature 
of potentials whose asymptotic tail is proportional to ${\rm e}^{-r/a}$ \cite{Ps}. 
This is because essential conclusions of our analysis will not change 
if the exact equalities involving $r$-dependence were replaced by asymptotic ones.
Whether redundant poles and zeros and the Heisenberg condition for other model cases,
including non-Hermitian scattering Hamiltonians \cite{SBK}, show similar behaviour
is the subject of future study. 

\acknowledgments

The work of AEM was supported by the Australian Research Council and UNSW Scientia Fellowship.

%%%%%%%%%%%%%%%%%%%%%%%%%%%%%%%%%%%%%%%%%%


\begin{thebibliography}{99}
\bibitem{Gmf}
$\Gamma(z)$ has no zeros, so the reciprocal gamma function 
$1/\Gamma(z)$ is a holomorphic function \cite{AS}.
\end{thebibliography}
\end{document}